\documentclass[12pt]{article}
\usepackage{epsfig}

\newlength{\abstractwidth}
\newcommand{\starttext}{
\setcounter{footnote}{0}
\renewcommand{\thefootnote}{\arabic{footnote}}}

\def\IR{\mathbb{R}}

\def\ID{\relax{\rm I\kern-.18em D}}
\def\IE{\relax{\rm I\kern-.18em E}}
\def\IF{\relax{\rm I\kern-.18em F}}
\def\IG{\relax\hbox{$\inbar\kern-.3em{\rm G}$}}
\def\IGa{\relax\hbox{${\rm I}\kern-.18em\Gamma$}}
\def\IH{\relax{\rm I\kern-.18em H}}
\def\II{\relax{\rm I\kern-.18em I}}
\def\IK{\relax{\rm I\kern-.18em K}}
\def\IP{\relax{\rm I\kern-.18em P}}
\def\I1{{\bf 1}}

\def\IR{\relax{\rm I\kern-.18em R}}
\def\OL#1{ \kern1pt\overline{\kern-1pt#1\kern-1pt}\kern1pt }

\begin{document}
\renewcommand{\theequation}{\thesection.\arabic{equation}}

\begin{titlepage}
\bigskip
\bigskip
\rightline{SU-ITP 02-01} \rightline{SLAC-PUB-9117}
\rightline{hep-th/0201016}

\bigskip\bigskip\bigskip\bigskip\bigskip

\centerline{\Large \bf {Higher-Dimensional Quantum Hall Effect }}
\bigskip
\centerline{\Large \bf { in String Theory}}
\bigskip\bigskip
\bigskip\bigskip
\centerline{ Michal Fabinger
 }
\medskip\medskip\medskip\medskip
\centerline{Department of Physics and SLAC} \centerline{Stanford \
University} \centerline{Stanford, CA  94305-4060}
\bigskip
\centerline{\tt fabinger@stanford.edu}
\medskip
\medskip
\bigskip\bigskip
\begin{abstract}

We construct a string theory realization of the $4+1$d quantum
Hall effect recently discovered by Zhang and Hu. The string theory
picture contains coincident D4-branes forming an $S^4$ and having
D0-branes (i.e.\ instantons) in their world-volume. The charged
particles are modelled as string ends. Their configuration space
approaches in the large $n$ limit a {$\bf CP^3$}, which is an
$S^2$ fibration over $S^4$, the extra $S^2$ being made out of the
Chan-Paton degrees of freedom. An alternative matrix theory
description involves the fuzzy $S^4$. We also find that there is a
hierarchy of quantum Hall effects in odd-dimensional spacetimes,
generalizing the known cases in $2+1$d and $4+1$d.

\medskip
\noindent
\end{abstract}

\end{titlepage}
\starttext \baselineskip=18pt
\setcounter{footnote}{0}
\setcounter{equation}{0}

\section{Introduction}

Recently, condensed matter physicists Zhang and Hu found a
remarkable theoretical construction of a new physical phenomenon
-- a 4+1 dimensional quantum Hall effect \cite{Zhang:2001xs} on an
$S^4$. In this note we will describe a simple brane construction
which reproduces the same physics.

In fact, there are many connections between string theory and the
ordinary $2+1$d quantum Hall physics which were found during the
last two years. The first of them was a construction of Bernevig,
Brodie, Susskind and Toumbas \cite{Brodie:2000yz} reproducing the
quantum Hall effect on a sphere (see fig.1), followed by a series
of papers including \cite{StringyQHE}. Another line of progress
(including \cite{Matrix-QHE}) was started by the proposal of
Susskind \cite{Susskind:2001fb} that the granular structure of the
quantum Hall fluid can be captured by making the ordinary
Chern-Simons description non-commutative. This model can also be
obtained from the Lagrangian of a charged particle moving in
magnetic field by replacing its coordinates by matrices
\cite{Susskind:2001fb}, in the spirit of matrix theory for
D0-branes.

The $4+1$d quantum Hall system of \cite{Zhang:2001xs} is the
dynamics of particles in a large representation of $SU(2)$ moving
under the influence of the homogeneous instanton of $SU(2)$ on an
$S^4$. An important point in making a connection of this system to
string theory is to translate the $SU(2)$ dynamics of
\cite{Zhang:2001xs} into the dynamics of fundamentals of $U(n)$
moving in a background $U(n)$ field.

The string theory  construction itself  is a close analog of
\cite{Brodie:2000yz} (see fig. 2). The four-sphere is modelled by
a stack of coincident spherical D4-branes with a homogeneous
instanton of $U(n)$ in their world-volume. This instanton has
instanton number $N=\frac{1}{6}(n-1)n(n+1)$, and can be also
thought of as $N$ D0-branes in the D4-brane world-volume. Note
that precisely this system was constructed also in matrix theory
\cite{Castelino:1997rv}, and is referred to as the `fuzzy
four-sphere'. The charged particles themselves can be modelled as
the ends of fundamental strings connecting the spherical D4-branes
to a stack of flat D4-branes placed at the center of the $S^4$.
This system by itself will not be stable, but we will  briefly
comment on possible ways of stabilizing it. Note also that the
question of the string end statistics is subtle. However, we can
expect logic similar to \cite{Brodie:2000yz} to apply here as
well.

The string theory point of view makes the analogy between the
ordinary $2+1$d and the $4+1$d quantum Hall systems rather close
-- one can be obtained from another simply by changing the
dimensionality of the D-branes. The QHE corresponds to the
dynamics of strings ending on either a fuzzy $S^2$ or a fuzzy
$S^4$.

This point of view also provides an insight into the question of
generalizing the QHE to dimensions higher than four. This can be
achieved simply by replacing the fuzzy $S^2$ or $S^4$ by a fuzzy
$S^6$ or $S^8$. It is probably very hard to stabilize such systems
in string theory. On the other hand, we can also think of them as
a simple quantum mechanics in a non-dynamical background field,
out of the framework of string theory, in which case such problems
do not arise. Then we can even talk about QHE on an $S^{2k}$ for
$k > 4$. The corresponding homogeneous background gauge field can
be found for example using the techniques of
\cite{Ramgoolam:2001zx,Ho:2001as}.

\section{Review: $2+1$d quantum Hall effect on an $S^2$}
\label{Review2d}

Here we briefly mention some basic properties of charged particles
moving in a constant magnetic field on $S^2$
\cite{Haldane:1983xm}. The magnetic field is sourced by a magnetic
monopole inside the $S^2$, such that the total magnetic  flux
through the sphere is $n-1 \equiv 2I \in {\bf Z}$, say $I>0$. The
single-particle Hamiltonian can be written as
\begin{eqnarray}
H=\frac{1}{2MR^2}\sum_{\mu<\nu}\Lambda_{\mu\nu}^2,
\label{H2}
\end{eqnarray}
where $\mu, \nu = 1,2,3$; $\  \Lambda_{\mu\nu}\equiv -i(x_\mu
D_\nu - x_\nu D_\mu)$; and $\tilde x_\mu \equiv R x_\mu$ are the
embedding coordinates of the $S^2$. This Hamiltonian is invariant
under $SO(3)$ transformations generated by $L_{\mu\nu}\equiv
\Lambda_{\mu \nu} + I \epsilon_{\mu \nu \rho} x_\rho$. The
spectrum follows from the relation
$\sum_{\mu<\nu}\Lambda_{\mu\nu}^2 = \sum_{\mu<\nu}L_{\mu\nu}^2 -
I(I+1)$ and from the fact that $\sum_{\mu<\nu}L_{\mu\nu}^2$ has
eigenvalues $l(l+1)$ with $l=I, I+1, I+2 \dots$

The lowest Landau level states transform in the $l=I$
representation of $SO(3)$, so their degeneracy is $n=2I+1$. Lowest
Landau states localized about a particular point $x_\mu$ on the
$S^2$ are eigenstates of $x_\mu I_\mu$, where $I_\mu$ are the
$l=I$ representation matrices of $SO(3)$. $I_\mu$ can be also
viewed as the uplift of 3d Euclidean Dirac gamma matrices (i.e.
the Pauli matrices) from the $l=\frac{1}{2}$ representation to the
$2I$-th
 symmetric tensor power of the $l=\frac{1}{2}$ representation
(i.e. to the $l=2I$ representation). In other words
\begin{eqnarray}
I_\mu \sim (\sigma_\mu \otimes 1 \otimes 1 \cdots \otimes 1 +
1\otimes \sigma_\mu  \otimes 1 \cdots \otimes 1 + \cdots +
 1\otimes 1 \otimes 1 \cdots \otimes  \sigma_\mu)_{sym}.
\end{eqnarray}
This point of view will be useful in section
\ref{FuzzyS4WithoutStringTheory}.

\subsection{The string theory picture}

There is a cool string theory realization of this kind of quantum
Hall effect due to Bernevig, Brodie, Susskind and Toumbas
\cite{Brodie:2000yz}. To construct their quantum Hall system (see
fig.1), one starts with a spherical D2-brane and dissolves $n-1$
D0-branes in it. Then one takes $K$ coincident D6-branes extended
in the directions perpendicular to the ones where the D2 lives,
and moves them to the center of the two-sphere. When the D6-branes
cross the D2's, the Hanany-Witten effect \cite{Hanany:1996ie}
produces fundamental strings stretching between them. This
configuration can be in equilibrium due to repulsion of D6-branes
and D0's. We know that D0-branes behave as magnetic flux in the
world-volume of D2's and that string ends are charged under the
world-volume gauge field. As a result, the low-energy physics is
essentially the one described in the previous paragraph.

The D2 with D0 brane charge can be also thought of as $n-1$
D0-branes expanded into a spherical configuration which has an
induced local D2-brane charge \cite{FuzzySphere}, provided the
separation between neighboring D0's is much smaller than the
string length, and $n\gg 1$. In this picture, we can think of the
strings as being connected to the constituent D0-branes. Assume
for simplicity that $K=1$. The lowest Landau levels should
correspond to unexcited stings, so their degeneracy should be
equal to $n-1$. This indeed agrees with the original picture in
the previous section, up to a subleading correction -- a unit
shift of $n$.

\smallskip
\centerline{\epsfxsize=0.3\textwidth \epsfbox{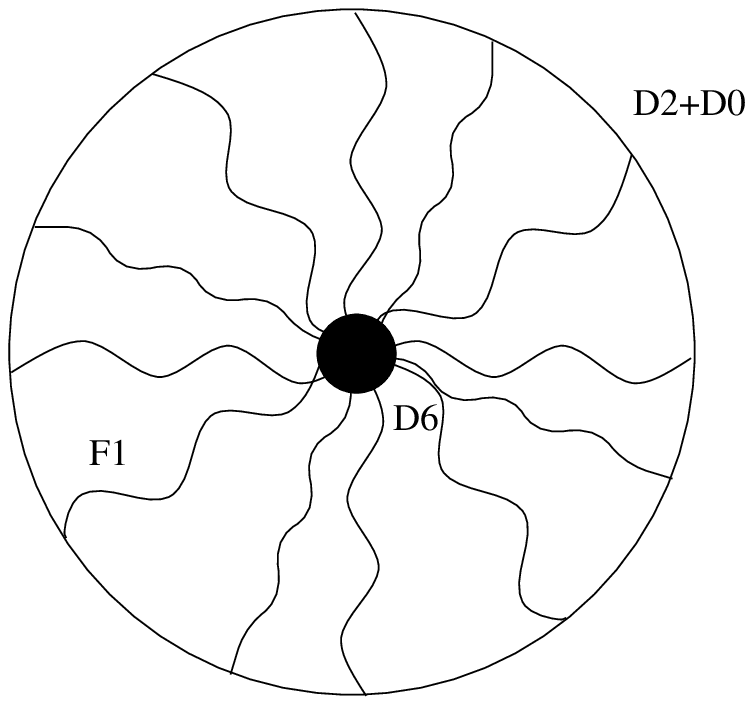}}
\smallskip
\centerline{Figure 1: This picture, stolen from
\cite{Brodie:2000yz}, shows
 a string theory realization }
 \centerline{of the
$2+1$d quantum Hall effect on a two-sphere.}
\smallskip

\vskip 5mm

\section{Review: $4+1$d quantum Hall effect on an $S^4$}

The 4+1d quantum Hall physics of \cite{Zhang:2001xs} is the
quantum mechanics of massive charged particles moving on a
four-sphere in a time-independent $SU(2)$ gauge field. This field
is taken to be the homogeneous $SU(2)$ instanton (i.e. a gauge
configuration of a non-vanishing second Chern class, with
instanton number one).

The charged particles are in some definite representation of the gauge group corresponding to some value $I \equiv (n-1)/2$ of the
$SU(2)$ `isospin', and can be described by $n$-component vectors. It was found in \cite{Zhang:2001xs} that eventually, one has to
take $n$ very large to obtain a reasonable thermodynamic limit.

\subsection{The second Hopf map}
\label{TheSecondHopfMap}
 To parameterize the four-sphere, we will use (slightly modified) conventions
of \cite{Zhang:2001xs}. The embedding coordinates  $\tilde x_\mu =
R x_\mu$ ($\mu=1..5$, $x_\mu x_\mu =1$) of the $S^4$ in a 5d
Euclidean space can be written as
\begin{eqnarray}
x_\mu = \bar\Psi_\alpha (\Gamma_\mu)_{\alpha\alpha'} \Psi_{\alpha'},
\ \ \, \ \ \
\bar\Psi_\alpha \Psi_\alpha =1.
\label{2ndHopf}
\end{eqnarray}
Here, the $\Gamma_\mu$ are $4 \times 4$ Euclidean Dirac gamma
matrices,  $\{\Gamma_\mu,\Gamma_\nu\}=2\delta_{\mu\nu}$, and
$\Psi_{\alpha}$ is a four component complex spinor, the
fundamental spinor of $SO(5)$. Obviously, there is a redundancy in
parameterizing the $S^4$ with $\Psi_{\alpha}$. Indeed, if we
choose the gamma matrices as
\begin{eqnarray}
& & \Gamma^{i} \!=\! \left( \begin{array}{cc}
               0   & i \sigma_i  \\
               -i \sigma_i       &  0  \end{array} \right), \ \  \ \ i=1..3, \ \
 \Gamma^4\!=\! \left( \begin{array}{cc}
               0          & 1  \\
               1          & 0            \end{array} \right), \ \  \ \
 \Gamma^5 \!=\! \left( \begin{array}{cc}
                1         & 0  \\
                0         & -1            \end{array} \right), \\
\end{eqnarray}
we can write $\Psi_{\alpha}$ in the following form,
\begin{eqnarray}
& &  \left( \begin{array}{c}
               \Psi_1    \\
               \Psi_2  \end{array} \right) =
               \sqrt{\frac{1+x_5}{2}}
               \left( \begin{array}{c}
               u_1    \\
               u_2  \end{array} \right), \ \  \
\left( \begin{array}{c}
               \Psi_3    \\
               \Psi_4  \end{array} \right) =
               \sqrt{\frac{1}{2(1+x_5)}} (x_4-ix_i \sigma_i)
               \left( \begin{array}{c}
               u_1    \\
               u_2  \end{array} \right),
\end{eqnarray}
with some two-component complex spinor $(u_1, u_2)$ satisfying
$\bar u_{\sigma} u_{\sigma} =1$. Varying $u_{\sigma}$ changes
$\Psi_{\alpha}$, but leaves $x_\mu$ invariant. The space of
different $u_{\sigma}$ is an $S^3$, $\Psi_{\alpha}$ span an $S^7$,
and as we said,  $x_\mu$ are coordinates on an $S^4$. This
construction of $S^7$ as a fibration of $S^3$ over $S^4$ is known
as the second Hopf map. If we mod out this $S^7$ by the $U(1)$
phase rotations of $\Psi_\alpha$, we obtain ${\bf CP^3}$ as an
$S^2$ fibration over $S^4$.

\subsection{The quantum Hall mechanics}

The gauge field of the homogeneous $SU(2)$ instanton on the $S^4$, i.e. of the Yang monopole \cite{Yang:1977qv},
can be written as
\begin{eqnarray}
A_\mu = \frac{a_\mu}{R} ;   \ \ a_A =
-\frac{i}{1+x_5}\eta_{iAB} x_B {\frac{\sigma_i}{2}}, \ \ a_5 = 0,   \label{potential}\\
\eta_{iAB}=\epsilon_{iAB4}+\delta_{iA}\delta_{4B}-
\delta_{iB}\delta_{4A}, \ \ A,B=1..4
\end{eqnarray}
where $\sigma_i$ are the Pauli matrices. (The gauge connection is $a_\mu
dx_\mu$,
with the restriction $x_\mu dx_\mu$ =0). This potential can also
be obtained as $\bar u_\sigma a_{\sigma \sigma'} u_{\sigma'}
=\bar\Psi_\alpha d\Psi_\alpha$ , which means that
 (\ref{potential}) is precisely what we get if we think of the $S^3$ in the second Hopf map as an $SU(2)$ gauge field living on
 the $S^4$.

The single-particle Hamiltonian  of \cite{Zhang:2001xs} is simply
the kinetic energy of the charged particles plus the interaction
with the $SU(2)$ gauge field,
\begin{eqnarray}
H=\frac{1}{2MR^2}\sum_{\mu<\nu}\Lambda_{\mu\nu}^2.
\label{H}
\end{eqnarray}
Here, $M$ is the mass of the particle, and $\Lambda_{\mu\nu}=
-i(x_\mu D_\nu - x_\nu D_\mu)$, with some appropriate covariant
derivatives $D_\mu$ whose form depends on the $SU(2)$
representation chosen. In general, $H$ will be an $n\times n$
matrix, with $I\equiv (n-1)/2$ being the value of the $SU(2)$ `isospin'.

The Hamiltonian (\ref{H}) is $SO(5)$ invariant, though the usual $SO(5)$ action has to be accompanied by an extra isospin
rotation. In other words, the $SO(5)$ generators are
\begin{eqnarray}
L_{\mu\nu} \equiv \Lambda_{\mu \nu} - i f_{\mu\nu},\quad
f_{\mu\nu}\equiv [D_\mu ,D_\nu].
\end{eqnarray}
 As a result,
the energy eigenstates form $SO(5)$ representations. Any $SO(5)$
representation can be labelled by two integers $r_1 \ge r_2\ge 0$,
the row lengths of the corresponding Young diagram.\footnote{A
nice and succinct review of $SO(2k+1)$ and $SO(2k)$
representations can be found in \cite{Ramgoolam:2001zx}.} The
dimension of the representation is
\begin{equation}
D(r_1, r_2) = \frac{1}{6}(r_1 + r_2 +2)(r_1 - r_2 +1)(3+2r_1)(1+2r_2).
\end{equation}
Expressed in terms of the variables of \cite{Zhang:2001xs},
\begin{equation}
p=r_1 + r_2, \quad q = r_1 - r_2,\quad p\ge q \ge 0,
\end{equation}
this becomes
\begin{equation}
\tilde D(p,q) =(1+q)(1+p-q)(1+\frac{p}{2})(1+\frac{p+q}{3}).
\end{equation}
The variables $p$ and $q$ satisfy \cite{Zhang:2001xs}
\begin{equation}
  \quad p-q=2I, \quad I\equiv \frac{n-1}{2},
\end{equation}
implying
\begin{equation}
 r_1 \ge r_2 = I.
\end{equation}
The reason why there is a restriction on the value of  $r_2$ is
that the particles are point-like, and therefore they cannot have
two independent `angular momenta', i.e. two independent charges
under the Cartan subalgebra of the $SO(5)$. The value of $r_2$ is
non-zero just because the generators of this $SO(5)$ are not
exactly equal to the dynamical angular momenta if  $I \neq 0$.

Using the identity
\begin{equation}
H = \frac{1}{2MR^2} \sum_{\mu<\nu}\Lambda_{\mu\nu}^2 =
\frac{1}{2MR^2} \left( \sum_{\mu<\nu}L_{\mu\nu}^2 -
2I(I+1)\right),
\end{equation}
one can  easily find the energy corresponding to a given $SO(5)$
representation,
\begin{equation}
E(I,q)=\frac{1}{2MR^2}(2I + (2I+3)q + q^2).
\end{equation}
We see that $q$ plays the role of the Landau level index. The
degeneracy of the lowest Landau level ($q=0$) is
\begin{equation}
\tilde D(n-1,0) = \frac{1}{6}n(n+1)(n+2).
\label{LLLDegeneracy}
\end{equation}

\section{$U(n)$ interpretation of the $4+1$d quantum Hall effect}

To make any connection to string theory, we have to get rid of the extremely high gauge representations of the charged particles.
This can be done very easily. Note that the gauge connection of the charged particles due to (\ref{potential}) can be obtained
from (\ref{potential}) by replacing $\frac{1}{2}\sigma_i$ with appropriate generators $I_i$ of the $SU(2)$ Lie algebra
\cite{Zhang:2001xs}, $[I_i,I_j]=i\epsilon_{ijk} I_k$,
\begin{eqnarray}
A_\mu = \frac{a_\mu}{R} ;   \ \ a_A = -\frac{i}{1+x_5}\eta_{iAB}
x_B {I_i}, \ \ a_5 = 0.
\label{potential2}
\end{eqnarray}
The particles are described by $n$-component vectors, and both
(\ref{H}) and (\ref{potential2}) are $n \times n$ matrices. Thus,
the system is equivalent to particles in the fundamental
representation of $U(n)$ moving under the influence of a $U(n)$
gauge field, given by (\ref{potential2}). The equivalence is
possible only because here we are treating all the gauge fields as
a fixed background. Of course, if they were dynamical, there would
be a big difference between $SU(2)$ and $U(n)$ gauge fields.

The gauge configuration (\ref{potential2}) is actually the homogeneous
instanton solution of $U(n)$ with instanton number
\begin{eqnarray}
N = \frac{1}{6}(n-1)n(n+1).
\end{eqnarray}
A review of these homogeneous instantons can be found for example
in Appendix B of \cite{Constable:2001ag}. From string theory we
know that for large $R$ we can think of this gauge field as $N$
D0-branes in the world-volume of  $n$ coincident  spherical
D4-branes.

\section{String theory construction of the $4+1$d quantum Hall effect}

Having identified the brane construction of the gauge field on the
four-sphere in the previous section, it is easy to model the full
system, including the `electrons'. This can be done in a way
analogous to \cite{Brodie:2000yz}. We start with $n$ coincident
spherical D4-branes and spread $\frac{1}{6}(n-1)n(n+1)$ D0-branes
in their world-volume. Say these D4-branes are extended in $\tilde
x_\mu, \mu=1..5$. Consider also $m$ flat infinite D4-branes far
from the four-sphere, extended in $\tilde x_\mu, \mu=6..9$. Now
move the $m$ D4-branes to the center of the four-sphere. The
Hanany-Witten effect \cite{Hanany:1996ie}
 produces $m n$ fundamental strings connecting the D4-branes at the center to the ones forming the four-sphere. The string ends
 are fundamentals of the D4-brane $U(n)$ gauge group, which is precisely what we need to interpret them as the `electrons' of
 \cite{Zhang:2001xs}.

Since the system breaks supersymmetry, it cannot be fully stable.
It is not even metastable, and it will immediately start to
collapse. Of course, if we want to study the stringy quantum Hall
physics of this system, we have to make it at least metastable or
very slowly decaying. One of the possible ways might be placing
the whole system into a fluxbrane \cite{fluxbranes} with a
non-zero Ramond-Ramond field strength $F_{\it 6}=dC_{\it 5}$ with
Lorentz $SO(5,1)$ and rotational $SO(4)$ symmetry \cite{Mark}. (A
similar set-up for two-spheres was studied in
\cite{Costa:2001if}.) If we make the RR field strong enough, the
spherical D4-branes want to expand. On the other hand, the field
strength of the RR field goes to zero at infinity, so one might
expect that the system will reach some equilibrium radius. Of
course, the stability with respect to non-spherical deformations
will be important, too. In this paper, we will leave this question
open.

\bigskip
\centerline{\epsfxsize=0.3\textwidth \epsfbox{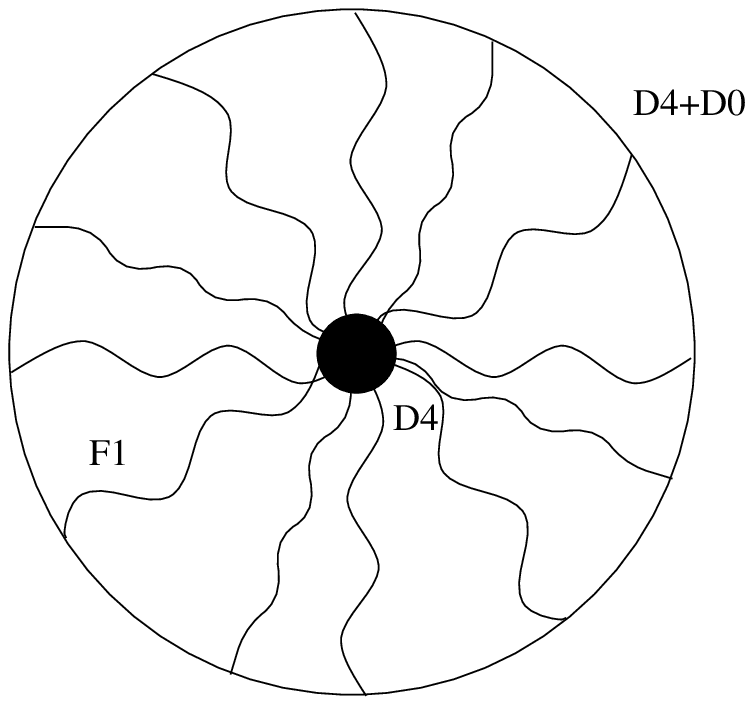}}
\smallskip
\centerline{Figure 2: The string theory picture of the $4+1$d
quantum Hall effect  on an $S^4$ can be}
 \centerline{ obtained from
the one at fig. 1 simply by changing the dimensionality of the
D-branes.}
\smallskip

\vskip 5mm

\subsection{Fuzzy four-sphere interpretation}

There is a nice interpretation of the spherical D4-branes carrying D0-brane charge --
the non-commutative (or fuzzy) four-sphere
\cite{Grosse:1996mz}. Such a system can be constructed
in matrix theory, where one starts with  $\frac{1}{6}(n-1)n(n+1)$ D0-branes and
expands them into an $S^4$ \cite{Castelino:1997rv}. The local D4-brane charge of such a configuration was found to be $n$.

As a result, the system can be described either in terms of
D4-branes with a D0-brane charge, or in terms of D0-branes
expanded into a fuzzy $S^4$. Just like in \cite{FuzzySphere},
these two pictures should agree at the leading order in $n$, but
there is no reason to expect that also the subleading corrections
are the same.

Now, we would like to see what implications the matrix fuzzy $S^4$
picture has for the quantum Hall physics. As in
\cite{Brodie:2000yz}, one can think of the fundamental string as
being connected to the D0-branes forming the $S^4$. Let us for
simplicity assume that there is just one D4 at the center of the
$S^4$ ( i.e. $m=1$). We can expect that there is one unexcited
string state for each D0-brane. Because unexcited strings have the
lowest possible energy, they should correspond to the lowest
Landau levels. As a result, we expect the lowest Landau level to
be  $\frac{1}{6}(n-1)n(n+1)$-fold degenerate. Indeed, at the
leading order in $n$, this agrees with (\ref{LLLDegeneracy}).

\subsection{The magic geometry of the fuzzy $S^4$}

We have seen that the $4+1$d QHE is in a good sense really
$4+1$-dimensional -- it is the physics of $U(n)$ fundamentals on
an $S^4$. However, if we kept $n$ large and asked the fundamentals
what configuration space they live in, they would describe it
(especially after reading section \ref{TheSecondHopfMap}) as an
$S^2$ fibration over $S^4$, the  $S^2$ fibre representing the
possible orientations of the $SU(2)$ `isospin'. They could also
mention that the space is actually a ${\bf CP^3}$, if they wanted
to be as precise as Edward Witten, who pointed out this
interesting fact. The { $\bf CP^3$} can be viewed\footnote{I would
like to thank Lubo\v s Motl for discussions about this point.} as
$Sp(2)/SU(2)\times U(1)$, where $SU(2)\times U(1) \subset
SU(2)\times SU(2) = Sp(1)\times Sp(1) \subset Sp(2)$.
Alternatively, it is also the bundle of unit anti-self-dual
two-forms on $S^4$.

This agrees with the matrix theory calculations of
\cite{Ho:2001as}, where it was found that the full matrix algebra
of the fuzzy $S^4$ approaches in the large $n$ limit the algebra
of functions on $SO(5)/U(2) =Sp(2)/SU(2)\times U(1)$.

Associated with these two points of view are two alternative
descriptions of the same physics -- a $6+1$-dimensional
\cite{Lenny} and a $4+1$-dimensional. This kind of duality was
discussed in \cite{Ho:2001as} in the context of ordinary field
theory on the fuzzy $S^4$. Probably the most interesting aspect of
this correspondence is that from the D4-brane point of view, the
two extra dimensions are made out of the D4-brane gauge
fields.\footnote{For later developments see \cite{NewPapers,
Bernevig:2002eq} }

\subsection{How to see the fuzzy $S^4$ without using string theory}
\label{FuzzyS4WithoutStringTheory}

The fuzzy four-sphere structure of the lowest Landau level ($q=0$,
$p=2I$) of \cite{Zhang:2001xs} can be seen quite easily even
without the machinery of string theory. The single particle states $\tilde
\Psi$ transform in the $(r_1=I, r_2 = I)$ of $SO(5)$, which is the
$2I$-th symmetric tensor power of the fundamental spinor
$\Psi$ of $SO(5)$, i.e. $(\frac{1}{2},\frac{1}{2})$. $\tilde \Psi$
can be thought of as a vector with
$P\equiv\frac{1}{6}(p+1)(p+2)(p+3)$ components, because the lowest Landau level is
$P$-fold degenerate.

In analogy to section \ref{Review2d}, the lowest Landau states
localized about a certain point $x_\mu$ on the $S^4$ are
eigenstates of $x_\mu\tilde X_\mu$. The operators $X_\mu$ can be
obtained by uplifting the gamma matrices $\Gamma_\mu$ from the
$(\frac{1}{2},\frac{1}{2})$ representation to
$(I,I)=(\frac{p}{2},\frac{p}{2})$, cf. also (\ref{2ndHopf}). In
other words,
\begin{eqnarray}
X_\mu \sim (\Gamma_\mu \otimes 1 \otimes 1 \cdots \otimes 1 +
1\otimes \Gamma_\mu  \otimes 1 \cdots \otimes 1 + \cdots +
 1\otimes 1 \otimes 1 \cdots \otimes  \Gamma_\mu)_{sym}.
 \label{FuzzyTildeX}
\end{eqnarray}
Moreover, since the wave-functions $\tilde \Psi$ transform in the
$(I,I)$ of $SO(5)$, we have
\begin{eqnarray}
L_{\mu\nu} \tilde \Psi = -\frac{i}{4} [ X_\mu,  X_\nu] \tilde
\Psi, \label{FuzzyL}
\end{eqnarray}
and
\begin{eqnarray}
H \tilde \Psi = \frac{1}{2MR^2}\! \left( \sum_{\mu < \nu} L^2_{\mu\nu}   - I(I+1)
  \right) \! \tilde \Psi= \frac{1}{2MR^2}\! \left( - \sum_{\mu < \nu} \frac{1}{16} \ \! [ X_\mu,  X_\nu]^2
  - I(I+1)  \right)\! \tilde \Psi. \label{FuzzyH}
\end{eqnarray}
Now, the connection to the fuzzy four-sphere became obvious. In
(\ref{FuzzyTildeX}) we recognize the matrices representing the
fuzzy $S^4$ \cite{Castelino:1997rv}, and (\ref{FuzzyL}) are the
generators of its rotations. However, it is hard to judge from the
present calculations whether there is any deeper reason why
(\ref{FuzzyH}) resembles the usual matrix theory Hamiltonian.

\section{Generalization to Higher Dimensions}

 We have seen that the QHE on 2d or 4d spherical
surfaces corresponds to the quantum mechanics of charged particles
moving in the gauge field of the fuzzy $S^2$ or $S^4$.
Higher-dimensional fuzzy spheres have been constructed as well, at
least in matrix theory. Physics of charged particles (or string
ends) moving in their gauge field naturally generalizes the $2+1$d
and $4+1$d QHE. Let us briefly mention what properties of these
quantum Hall systems can be inferred from the matrix theory
results of \cite{Ramgoolam:2001zx,Ho:2001as}.

The fuzzy $S^6$ consists of
\begin{eqnarray}
N_6 = \frac{1}{360}(n+1)(n+2)(n+3)^2(n+4)(n+5)
\end{eqnarray}
D0-branes and its D6-brane charge is
\begin{eqnarray}
D_6 = \frac{1}{6}(n+1)(n+2)(n+3).
\end{eqnarray}
Therefore the $6+1$d QHE should be the dynamics of the
fundamentals of $U(D_6)$ in a homogeneous background $U(D_6)$
gauge field with a non-zero third Chern class. The degeneracy of
the lowest Landau level should be $N_6$, up to subleading shifts of $n$.
 The case of the fuzzy
$S^8$ is analogous, with
\begin{eqnarray}
N_8 =
\frac{1}{302400}(n+1)^2(n+2)^2(n+3)^4(n+4)^4(n+5)^4(n+6)^2(n+7)^2,
\end{eqnarray}
and
\begin{eqnarray}
D_8 = \frac{1}{360}(n+1)(n+2)(n+3)^2(n+4)(n+5).
\end{eqnarray}
In general for a fuzzy $S^{2k}$, the configuration space of the
fundamentals should closely approximate $SO(2k+1)/U(k)$ at large
$n$.

\section{Conclusions}

 We have described a string theory system reproducing
the quantum Hall effect of Zhang and Hu. Moreover, we have seen
that there is an extremely close relationship between this $4+1$d
and the ordinary spherical $2+1$d quantum Hall system -- they are
the first two elements in a hierarchy of quantum Hall systems on
even-dimensional spheres.

Of course, many questions related to the $4+1$d quantum Hall
effect are still open -- for example the issue of finding an
appropriate non-commutative Chern-Simons description\footnote{This
problem has now been solved \cite{Bernevig:2002eq}.}, or
understanding well the physics of the edge excitations. It is
natural to expect that string theory will play an important role
in answering these questions, judging from the many recently
discovered connections between string theory and the ordinary
quantum Hall effect.

\vskip 5mm {\bf Acknowledgements.} I would like to thank
B.~Freivogel, J.~McGreevy, S.~Hellerman, J.~P.~Hu, M.~Kleban,
X.~Liu, L.~McAllister, L.~Motl, M. Van Raamsdonk, S.~Shenker,
S.~C.~Zhang, and especially L.~Susskind for extremely helpful
discussions. This work was supported in part by a Stanford
Graduate Fellowship, by NSF grant PHY-9870115, and by the
Institute of Physics, Academy of Sciences of the Czech Republic
under grant no. GA-AV{\v C}R~A10100711.

\end{document}